\documentclass[12pt]{article} 
\usepackage{simpleConference}
\usepackage{times}
\usepackage{graphicx}
\usepackage{amssymb}
\usepackage{url,hyperref}

\usepackage{xcolor}
\usepackage{listings}
\usepackage{caption}
\usepackage{empheq}
\usepackage{float}

\begin{document}
\title{SymFields: An Open Source Symbolic Fields Analysis Tool for General Curvilinear Coordinates in Python}
\author{CHU Nan \\
\\
Institute of Plasma Physics, Chinese Academy of Sciences, Hefei 230031, China \\
\today
\\
\\
chunan@ipp.ac.cn \\
}
\maketitle
\thispagestyle{empty}

\abstract{An open source symbolic tool for vector fields analysis 'SymFields' is developed in Python. The SymFields module is constructed upon Python symbolic module sympy, which could only conduct scaler field analysis. With SymFields module, you can conduct vector analysis for general curvilinear coordinates regardless whether it is orthogonal or not. In SymFields, the differential operators based on metric tensor are normalized to real physical values, which means your can use real physical value of the vector fields as inputs. This could greatly free the physicists from the tedious calculation under complicated coordinates.}

\section{Introduction}
The plasma physics MHD theory is a combination of fluid equations and Maxwell's equations. The MHD equations involve the solution to multiple vector fields: electric field $\vec{E}(\vec{R})$, magnetic field $\vec{B}(\vec{R})$, displacement vector $\vec{\xi}(\vec{R})$, etc. Thus, the derivation in MHD theory usually becomes very complicated not because of physics issue but due to tremendous long terms in equations. To avoid this kind of tedious work, symbolic calculation is developed in many programming languages. For example, the General Vector Analysis (GVA) toolbox for Mathematica software developed by Prof. H. Qin, which could conduct symbolic fields analysis for general coordinates with elegant expression after simplification \cite{Qin_1999_CPC}. Matlab and Python also have some elementary symbolic calculation functions. However, Mathematica and Matlab are commercial software with high price, which is hardly affordable to me. In open source Python language, there is already a fundamental symbolic calculation module sympy. However, it does not provide the general analysis functions for vector fields. The SageMath project is another useful open source symbolic calculation software (\url{https://www.sagemath.org/}). It provides many useful functions for symbolic calculation in vector fields. Although it depends on many modules in python, its source codes is not compatible with Python.
Since we do not wish to get involved with a new language, we would like to develop an open source symbolic fields analysis tool solely within Python. Therefore, we developed the SymFields module in Python for fields analysis in general coordinates. It could not only deal with the vector analysis in commonly seeing orthogonal coordinates, but also capable to analyse fields in non-orthogonal coordinates. In this paper, the second section talks about fields analysis related mathematics for general coordinates (orthogonal and non-orthogonal). The third section discusses the realization of field analysis symbolic calculation in SymFields module in python. The fourth section give benchmark examples to use SymFields in both orthogonal and non-orthogonal coordinates. The last section is summary. The SymFields module is available on github at (\url{https://github.com/DocNan/SymFields}) under GNU General Public License v3.0.

\section{Mathematics of general coordinates}
The multiple choose of coordinates in Mathematics can sometime make the formulas in fields analysis complicated. In orthogonal coordinates, the vector analysis can be simplified due to the orthogonality. The differential field operators can be easily expressed with Lame coefficients in orthogonal coordinates \cite{USTC_2008_book}. However, for the more general non-orthogonal coordinates, one must use metric tensor to express these operators. Therefore, we shall start our discussion from the general coordinates.

\subsection{Basic definitions in general curvilinear coordinates}
\subsubsection{Covariant and contra-variant vector basis}
In general $R^3$ curvilinear coordinates, we can define two sets of basis vectors. First we pick a point P in Cartesian coordinate with location vector: $\vec{R} = (x, y, z) = x\vec{e_x} + y\vec{e_y} + z\vec{e_z}$. Suppose for the same point P in general curvilinear coordinate, its coordinate is: $(\xi^1, \xi^2, \xi^3)$, where $\xi^i(\vec{R}) = \xi^i(x, y, z)$ is a function of the Cartesian coordinates. The contra-variant vector basis is defined as:
\begin{equation}
    \begin{cases}
        \vec{g^1} = \nabla\xi^1 \\
        \vec{g^2} = \nabla\xi^2 \\
        \vec{g^3} = \nabla\xi^3 \\
    \end{cases}
\end{equation}

The reciprocal covariant basis vector is defined as:
\begin{equation}
    \begin{cases}
        \vec{g_1} = \frac{\partial\vec{R}}{\partial\xi^1} \\
        \vec{g_2} = \frac{\partial\vec{R}}{\partial\xi^2} \\
        \vec{g_3} = \frac{\partial\vec{R}}{\partial\xi^3}
    \end{cases}
\end{equation}

They can be converted through the simple definition that:
\begin{equation}
    \vec{g_i} = \frac{\vec{g^j}\times\vec{g^k}}{\vec{g^i}\cdot(\vec{g^j}\times\vec{g^k})}, \ 
    \vec{g^i} = \frac{\vec{g_j}\times\vec{g_k}}{\vec{g_i}\cdot(\vec{g_j}\times\vec{g_k})}
\end{equation}

With the above definition we can easily find the orthogonality between the vector basis as: 
\begin{equation}
    \vec{g^i}\cdot\vec{g_j} = \delta^i_j = 
    \begin{cases}
        0\ (i \neq j) \\
        1\ (i = j)
    \end{cases}
\end{equation}

With the two set of basis vectors, any vector filed $\vec{A} = \vec{A}(\vec{R})$ can also be expressed in two equal ways:
\begin{equation}
    \vec{A} = \sum_i A_i\vec{g^j} = \sum_i A^i\vec{g_i}
\end{equation}

But we must pay attention that unlike the Cartesian coordinates, the basis vector of general curvilinear coordinates: $\vec{g_i}$ and $\vec{g^j}$ are not unit vector: $|\vec{g_i}| \neq 1 \neq |\vec{g^j}|$. Thus the vector components $A_i$ and $A^i$ under contra and covariant vector basis are not real physical value, instead they are called the contra- and covariant components of this vector field \cite{Piercey_lecture}.

\subsubsection{Contra- and covariant metric tensors}
The contra variant metric tensor is defined as: 
\begin{equation}
    g^{ij} = \vec{g^i}\cdot\vec{g^j} =
    \left (
    \begin{array}{ccc}
        \vec{g^1}\cdot\vec{g^1} &  \vec{g^1}\cdot\vec{g^2} &  \vec{g^1}\cdot\vec{g^3} \\
        \vec{g^2}\cdot\vec{g^1} &  \vec{g^2}\cdot\vec{g^2} &  \vec{g^2}\cdot\vec{g^3} \\
        \vec{g^3}\cdot\vec{g^1} &  \vec{g^3}\cdot\vec{g^2} &  \vec{g^3}\cdot\vec{g^3}
    \end{array}
    \right )
    = 
    \left (
    \begin{array}{ccc}
    g^{11} &  g^{12} & g^{13} \\
    g^{21} &  g^{22} & g^{23} \\
    g^{31} &  g^{32} & g^{33}
    \end{array}
\right )
\end{equation}

Similarly, the covariant metric tensor is defined as: $g_{ij} = \vec{g_i}\cdot\vec{g_j}$. Due to the position exchange property of dot product between vectors ($\vec{g_i}\cdot\vec{g_j} = \vec{g_j}\cdot\vec{g_i}$), we can easily find that the transposition of the metric tensor equals to itself as: $g_{ij} = g_{ji} = g_{ij}^T$. The contra and covariant basis vector share one import relation as the product of the two metric tensors is unit matrix: $g^{ij}g_{ij} = I$.

\subsubsection{Jacobian for the coordinates}
The Jacobian of a coordinate represent the element volume under this coordinate, it is defined as: 
\begin{equation}
    J = \vec{g_1}\cdot(\vec{g_2}\times\vec{g_3}) = \sqrt{|g_{ij}|} = \frac{\partial(x, y, z)}{\partial(\xi^1, \xi^2, \xi^3)} = \frac{\partial\vec{R}}{\partial\xi^1}\cdot(\frac{\partial\vec{R}}{\partial\xi^2}\times\frac{\partial\vec{R}}{\partial\xi^3})
\end{equation}

Due to the reciprocal relation between contra and covariant basis vectors, their Jacobian also has relationship as:
\begin{equation}
   J' = \vec{g^1}\cdot(\vec{g^2}\times\vec{g^3}) = \sqrt{|g^{ij}|} = \frac{1}{\sqrt{|g_{ij}|}} = \frac{1}{J} = \frac{\partial(\xi^1, \xi^2, \xi^3)}{\partial(x, y, z)} = \nabla\xi^1\cdot(\nabla\xi^2\times\nabla\xi^3)
\end{equation}

\subsection{Differential operators in general coordinates}
\subsubsection{Expression of differential operators with metric tensor}
Suppose we have a scaler field: $U = U(\xi^1, \xi^2, \xi^3)$ and a vector field: $\vec{A} = A^1\vec{g_1} + A^2\vec{g_2} + A^3\vec{g_3}$, where the contra-variant component $A^j = A^j(\xi^1, \xi^2, \xi^3)$ is a scaler field, then the differential operators in general curvilinear coordinates are \cite{Piercey_lecture, William_1991_book}:

* Nabla operator:
\begin{equation}
    \nabla \sim \frac{\partial}{\partial\xi^i}\nabla\xi^i = \vec{g^i}\frac{\partial}{\partial\xi^i}
\end{equation}

* Gradient:
\begin{equation}
    \nabla U = \sum_i \frac{\partial U}{\partial \xi^i}\nabla \xi^i = \sum_i \frac{\partial U}{\partial \xi^i}\vec{g^i} = \sum_{i, j} g^{ij}\frac{\partial U}{\partial\xi^i}\vec{g_j}
\end{equation} 

* Divergence:
\begin{equation}
    \nabla\cdot\vec{A} = \sum_i \frac{1}{J}\frac{\partial}{\partial\xi^i}(J A^i)
    \label{eq_divergence_using_metric}
\end{equation}

* Curl: 
\begin{equation}
    \nabla\times\vec{A} = 
    \frac{1}{J}\left |
    \begin{array}{ccc}
        \vec{g_1} & \vec{g_2} & \vec{g_3} \\
        \frac{\partial}{\partial\xi^1} & \frac{\partial}{\partial\xi^2} & \frac{\partial}{\partial\xi^3} \\
        A_1 & A_2 & A_3
    \end{array}
    \right |
    = \frac{1}{J}\epsilon_{ijk}\frac{\partial A_k}{\partial \xi^j}\vec{g_i}
\end{equation}

* Laplacian: 
Since Laplacian is the combination of gradient and divergence, and can be calculated directly with $\Delta \vec{A} = \nabla^2 \vec{A} = \nabla\cdot\nabla U$, we will not give the specific expression to it here.

\subsubsection{Normalization of differential operators in physical value unit}
A type of common error frequently occurs in using the differential operators with metric tensor. For an example, in cylinder coordinate, the Jacobian is: $J = H_r H_{\phi} H_z = r^2$. According to formula (\ref{eq_divergence_using_metric}), the divergence is calculated via: 
\begin{equation}
\begin{array}{ll}
    \nabla\cdot\vec{A} & = \sum_i \frac{1}{J}\frac{\partial}{\partial\xi^i}(JA^i) \\
    & = \frac{1}{r^2}(\frac{\partial}{\partial r}(r^2 A_r) + \frac{\partial}{\partial \theta}(r^2 A_{\phi}) + \frac{\partial}{\partial z}(r^2 A_z)) \\
    & = \frac{1}{r^2}(r^2 A_r) + \frac{\partial A_{\phi}}{\partial \phi} + \frac{\partial A_z}{\partial z}\ {\color{red} (Wrong)}
\end{array}
\end{equation}

However, this expression for divergence operator in cylinder coordinate is wrong, the correct one shall be: $\nabla\cdot\vec{A} = \frac{1}{r}\frac{\partial}{\partial r}(rA_r) + \frac{1}{r}\frac{\partial A_{\phi}}{\partial \phi} + \frac{\partial A_z}{\partial z}$. The reason that cause this error is that the differential operators given in the above section use non-unity basis vectors. If we want to really use them in physics, we need to make normalization and convert the basis vectors to unity vectors to get the real meaningful physical components values. Which will be critical for the realization of the SymFields we will be talking about in the next section. Suppose the unity contra and covariant basis vector as: $\vec{e_i} = \frac{\vec{g_i}}{|\vec{g_i}|} = \frac{\vec{g_i}}{H_i}$, $\vec{e^i} = \frac{\vec{g^i}}{|\vec{g^i}|} = H_i cos\theta_i\vec{g^i}$. Where we uses the relation that:
\begin{equation}
\vec{g^i}\cdot\vec{g_i} = \delta_i^i = 1 \Leftrightarrow |\vec{g^i}||\vec{g_i}|cos\theta_i = |\vec{g^i}| H_i cos\theta_i = 1
\end{equation}

Among them $H_i = |\vec{g_i}| = |\frac{\partial\vec{R}}{\partial\xi^i}|$ is the Lame coefficient in $\xi^i$ coordinate direction \cite{USTC_2008_book}, $\theta_i$ is the angle between the vectors $\vec{g_i}$ and $\vec{g^i}$. Here we make the physical value normalization to the differential operators with unity covariant basis vector $\vec{e_i}$.

Suppose we have a vector field: $\vec{A} = \sum_i A^i\vec{g_i} = \sum_i A_i\vec{g^i}$, its normalization shall be: $\vec{A} = \sum_i \bar{A^i}\vec{e_i} = \sum_i \bar{A_i}\vec{e^i}$, where: $\bar{A^i} = A^i H_i$, $\bar{A_i} = A_i|\vec{g^i}|$. Thus the contra and covariant components of the field function for differential operator can be easily replaced with physical value as:
\begin{equation}
    \begin{cases}
        A^i = \frac{\bar{A^i}}{H_i} \\
        A_i = \frac{\bar{A_i}}{|\vec{g^i}|} = \frac{\bar{A_i}}{|\nabla\xi_i|} = \bar{A_i}H_i cos\theta_i
    \end{cases}
\end{equation}

* Dot product normalization:
\begin{equation}
    \vec{A}\cdot\vec{B} = \sum_{ij}(\bar{A^i}\vec{e_i}) \cdot (\bar{B^j}\vec{e_j}) = \sum_{ij}\bar{A^i}\bar{B^j}\vec{e_i}\cdot\vec{e_j} = \sum_{ij}\bar{A^i}\bar{B^j}\frac{g_{ij}}{H_i H_j}
\end{equation}

Where $\vec{e_i}\cdot \vec{e_j} = \frac{\vec{g_i}}{H_i} \cdot \frac{\vec{g_j}}{H_j} = \frac{g_{ij}}{H_i H_j} = e_{ij} = cos\theta_{ij}$, $\theta_{ij}$ is the angle between covariant basis vectors $\vec{g_i}$ and $\vec{g_j}$.

* Cross product normalization:
\begin{equation}
    \begin{array}{ll}
    \vec{A}\times\vec{B} & = \sum_{ij}A_i\vec{g^i}\times B_j\vec{g^j} = \sum_{ijk} A_i B_j\frac{\epsilon_{ijk}}{J}\vec{g_k} = \sum_{ijk}A_i B_j \frac{\epsilon_{ijk}}{J}H_k\vec{e_k} \\
    & = \frac{1}{J}
        \left |
        \begin{array}{ccc}
            \vec{g_1} & \vec{g_2} & \vec{g_3} \\
            A_1       & A_2       & A_3       \\
            B_1       & B_2       & B_3          
        \end{array}
        \right |
      = \frac{1}{J}
        \left |
        \begin{array}{ccc}
            H_1\vec{e_1}        & H_2\vec{e_2}        & H_3\vec{e_3}        \\
            g_{1i}\bar{A^i}/H_i & g_{2i}\bar{A^i}/H_i & g_{3i}\bar{A^i}/H_i \\
            g_{1i}\bar{B^j}/H_j & g_{3j}\bar{B^j}/H_j & g_{3j}\bar{B^j}/H_j  
        \end{array}
        \right |  
    \end{array}
\end{equation}

Where $\epsilon_{ijk}$ is Levi–Civita symbol and $A_i = g_{ij}A^j$ \cite{Lebedev_2010_book}.

* Gradient normalization:
\begin{equation}
    \nabla U(\xi^1, \xi^2, \xi^3) = \sum_{i,j} g^{ij}\frac{\partial U}{\partial \xi^i}\vec{g_j} = \sum_{i,j} g^{ij}\frac{\partial U}{\partial \xi^i}H_j\vec{e_j}
\end{equation}

Notice that for scaler field function, $U = U(\vec{\xi})$ is already a normalized physical value. Thus we only need to convert the basis vectors of gradient to unity ones.

* Divergence normalization:
\begin{equation}
    \nabla\cdot\vec{A} = \sum_i \frac{1}{J}\frac{\partial}{\partial\xi^i}(JA^i) = \sum_i \frac{1}{J}\frac{\partial}{\partial\xi^i}(\frac{J\bar{A^i}}{H_i})    
\end{equation}

* Curl normalization:
\begin{equation}
\nabla\times\vec{A} = 
\frac{1}{J}\left |
\begin{array}{ccc}
\vec{g_1} & \vec{g_2} & \vec{g_3} \\
\frac{\partial}{\partial\xi^1} & \frac{\partial}{\partial\xi^2} & \frac{\partial}{\partial\xi^3} \\
A_1 & A_2 & A_3
\end{array}
\right |
= 
\frac{1}{J}\left |
\begin{array}{ccc}
H_1\vec{e_1} & H_2\vec{e_2} & H_3\vec{e_3} \\
\frac{\partial}{\partial\xi^1} & \frac{\partial}{\partial\xi^2} & \frac{\partial}{\partial\xi^3} \\
g_{1i}\bar{A^i}/H_i & g_{2i}\bar{A^i}/H_i & g_{3i}\bar{A^i}/H_i
\end{array}
\right |
\end{equation}
Where $A_i = g_{ij}A^j = g_{ij}\bar{A^j}/H_j$, also notice that Einstein's summation assumption is used here.

\subsubsection{Special case: differential operators in orthogonal coordinates}
In orthogonal coordinates, the Jacobian is the product of all Lame coefficients: $J = H_1 H_2 H_3$. The contra-variant metric tensor shall be:
\begin{equation}
    (g^{ij}) = 
    \left (
    \begin{array}{ccc}
        1/H_1^2 & 0       & 0 \\
        0       & 1/H_2^2 & 0 \\
        0       & 0       & 1/H_3^2
    \end{array}
    \right )
    = (\frac{\delta^{ij}}{H_i H_j})
\end{equation}
Where $\delta^{ij}$ is delta function, when $i = j$, $\delta^{ii} = 1$, otherwise it takes 0 value. And the angle between $\vec{g_i}$ and $\vec{g^i}$ is: 
\begin{equation}
\theta_i = 0 \Leftrightarrow \vec{g_i} \parallel \vec{g^i} \Leftrightarrow cos\theta_i = 1 \Leftrightarrow H_i |\vec{g^i}| = 1
\end{equation}

Thus the general normalized differential operators can be rewritten as:

* Gradient:
\begin{equation}
\nabla U = \sum_{i,j} g^{ij}\frac{\partial U}{\partial \xi^i}H_i\vec{e_i} = \sum_i \frac{\partial U/\partial\xi^i}{H_i}\vec{e_i}
\end{equation}

* Divergence:
\begin{equation}
\nabla\cdot\vec{A} = \sum_i \frac{1}{J}\frac{\partial}{\partial\xi^i}(\frac{J\bar{A^i}}{H_i}) = \frac{1}{H_1 H_2 H_3}\sum_i\frac{\partial}{\partial\xi^i}(\bar{A^i}H_j H_k)
\end{equation}

* Curl:
\begin{equation}
\nabla\times\vec{A} = 
\frac{1}{J}
\left |
\begin{array}{ccc}
H_1\vec{e_1} & H_2\vec{e_2} & H_3\vec{3_3} \\
\frac{\partial}{\partial\xi^1} & \frac{\partial}{\partial\xi^2} & \frac{\partial}{\partial\xi^3} \\
g_{1i}\bar{A^i}/H_i & g_{2i}\bar{A^i}/H_i & g_{3i}\bar{A^i}/H_i
\end{array}
\right |
= 
\frac{1}{H_1 H_2 H_3}
\left |
\begin{array}{ccc}
H_1\vec{e_1} & H_2\vec{e_2} & H_3\vec{e_3} \\
\frac{\partial}{\partial\xi^1} & \frac{\partial}{\partial\xi^2} & \frac{\partial}{\partial\xi^3} \\
H_1\bar{A^1} & H_2\bar{A^2} & H_3\bar{A^3}
\end{array}
\right |
\end{equation}

The above expression of differential operators in orthogonal coordinates are the same as the ones in textbooks \cite{USTC_2008_book}. This also proves the correctness of our normalization.

\section{Realization of symbolic calculation in Python}
To make the calculation inputs in a minimum state, we only need to set the coordinates $(\xi^1, \xi^2, \xi^3)$ and contra-variant metric tensor $g^{ij}$ as inputs for Python symbolic calculation. The related Lame coefficients and covariant metric tensor could all be calculate from the two inputs. 

\subsection{Calculation of covariant metric tensor}
Since the covariant metric tensor is the inverse matrix of contra-variant metric tensor. It is very easy to use sympy module to calculate the inverse matrix with Gaussian elimination method: $g_{ij} = (g^{ij})^{-1}$.

\subsection{Calculation of Lame coefficients}
Whether the coordinates is orthogonal or not, we can always get the Lame coefficients from the covariant metric tensor as: $H_i = |\vec{g_i}| = \sqrt{g_{ii}}$ and similarly we have: $|\vec{g^i}| = |\nabla\xi^i| = \sqrt{g^{ii}}$.

\subsection{Calculation of normalized Contra-variant components}
For the cross product and curl operator, if we set the vector field components as: $\vec{A} = A_i \vec{g^i}$. Then after calculation we can get the results with covariant basis as: $\vec{g_j}$. However, to make an agreement during normalization, we choose the contra-variant components of field $\bar{A^j}$ by default. Thus we have to convert the covariant components $A_i$ to normalized contra-variant components. This could also be achieved with the help of co-variant metric tensor as: $A_i = g_{ij}A^j = g_{ij}\bar{A^j}/H_j$.

\subsection{Functions in SymFields module}
Following the formulas in section 2, we developed several functions in SymFields module to realize the symbolic field analysis. They are:
{\footnotesize
\begin{lstlisting}
Metric()   # calculate contra- or covariant metric tensor from curvilinear coordinates 
Jacobian() # calculate Jacobian with given metric tensor
Lame()     # calculate Lame coefficients from given metric tensor
Dot()      # calculate dot product of two vectors
Cross()    # calculate cross product of two vectors
Grad()     # calculate gradient from given scalar field function
Div()      # calculate divergence from given vector field function 
Curl()     # calculate curl from given vector field function
\end{lstlisting}
}

After import * from SymFields module, you can use the above functions to realize fields analysis for general curvilinear coordinates, regardless whether it is orthogonal or not.

\section{Benchmark of vector field analysis with SymFields module}
\subsection{Benchmark of differential operators in cylinder coordinates}
\subsubsection{Curl of gradient}
In any curvilinear coordinates, the curl of gradient for a scaler field will always be zero: $\nabla\times(\nabla U) = 0$. Thus we can use this rule to test our code. The benchmark test code is presented here:

{\footnotesize
\begin{lstlisting}
import sympy
from SymFields import *

# cylinder coordinates
r, phi, z = sympy.symbols('r, phi, z')
X = [r, phi, z]
U = sympy.Function('U')
U = U(r, phi, z)

grad = Grad(U, X, coordinate='Cylinder')
curl_grad = Curl(grad, X, coordinate='Cylinder')
\end{lstlisting}
}

The output results are:
{\footnotesize
\begin{lstlisting}
In: grad
Out: 
\end{lstlisting}
}

\begin{figure}[H]
    \centering
    \includegraphics[height=0.045\linewidth]{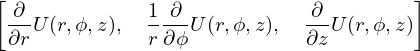}
\end{figure}

{\footnotesize
\begin{lstlisting}
In: curl_grad
Out: 
\end{lstlisting}
}

\begin{figure}[H]
    \centering
    \includegraphics[height=0.025\linewidth]{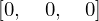}
\end{figure}

\subsubsection{Divergence of curl}
Similarly, the divergence of curl operator for a vector field will also be zero: $\nabla\cdot(\nabla\times\vec{A}) = 0$. The related code is presented here:
{\footnotesize
\begin{lstlisting}
A_r = sympy.Function('A_r')
A_phi = sympy.Function('A_phi')
A_z = sympy.Function('A_z')
A_r = A_r(r, phi, z)
A_phi = A_phi(r, phi, z)
A_z = A_z(r, phi, z)
A = [A_r, A_phi, A_z]

curl = Curl(A, X, coordinate='Cylinder')
div_curl = Div(curl, X, coordinate='Cylinder')
div_curl.doit()
\end{lstlisting}
}

{\footnotesize
\begin{lstlisting}
In: curl
Out: 
\end{lstlisting}
}

\begin{figure}[H]
    \centering
    \includegraphics[height=0.045\linewidth]{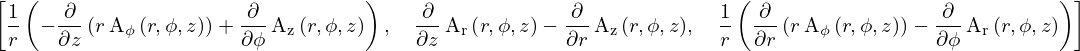}
\end{figure}

{\footnotesize
\begin{lstlisting}
In: div_curl.doit()
Out: 
\end{lstlisting}
}

\begin{figure}[H]
    \centering
    \includegraphics[height=0.015\linewidth]{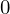}
\end{figure}

\subsubsection{Expression of Laplacian operator}
We also calculated the Laplacian operator with: $\nabla^2 U = \nabla\cdot\nabla U$. The detailed codes goes below:
{\footnotesize
\begin{lstlisting}
Laplacian = Div(grad, X, coordinate='Cylinder', evaluation=1)
\end{lstlisting}
}

{\footnotesize
\begin{lstlisting}
In: Laplacian
Out: 
\end{lstlisting}
}

\begin{figure}[H]
    \centering
    \includegraphics[height=0.045\linewidth]{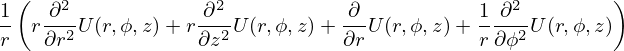}
\end{figure}
After manually simplification, you will find it is the correct expression of Laplacian operator in cylinder coordinate as the formula in textbooks.

\subsubsection{Build-in coordinates in SymFields module}
\begin{figure}[H]
    \centering
    \includegraphics[width=1.0\linewidth]{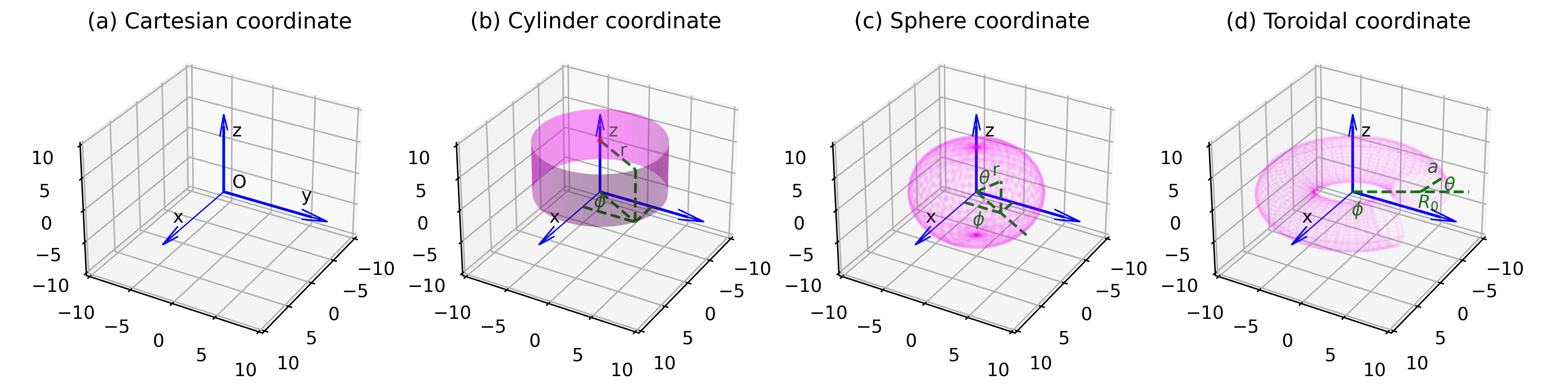}
    \caption{Build-in coordinates in SymFields module. From left to right, Cartesian, Cylinder coordinate, Sphere and Toroidal coordinates.}
\end{figure}
In SymFields module, we have already constructed the metric tensors for several frequently used orthogonal coordinates. They are: Cartesian, Cylinder, Sphere and Toroidal coordinates as shown in Fig. . To use these build in coordinates, you just need to set string value in the optional input like this: functionA(coordinate='Cylinder'). For other coordinates, you can first get the mapping function relation between this curvilinear coordinates ($\xi^1, \xi^2, \xi^3$) and Cartesian coordinates (x, y, z). Then you can calculate the related metric with function Metric() in SymFields module. Finally, you can conduct all the rest differential operations with the calculated metric tensor as inputs for the operator functions in SymFields module.

\subsection{Benchmark of differential operators in non-orthogonal coordinates}
\begin{figure}[H]
    \centering
    \includegraphics[height=0.3\linewidth]{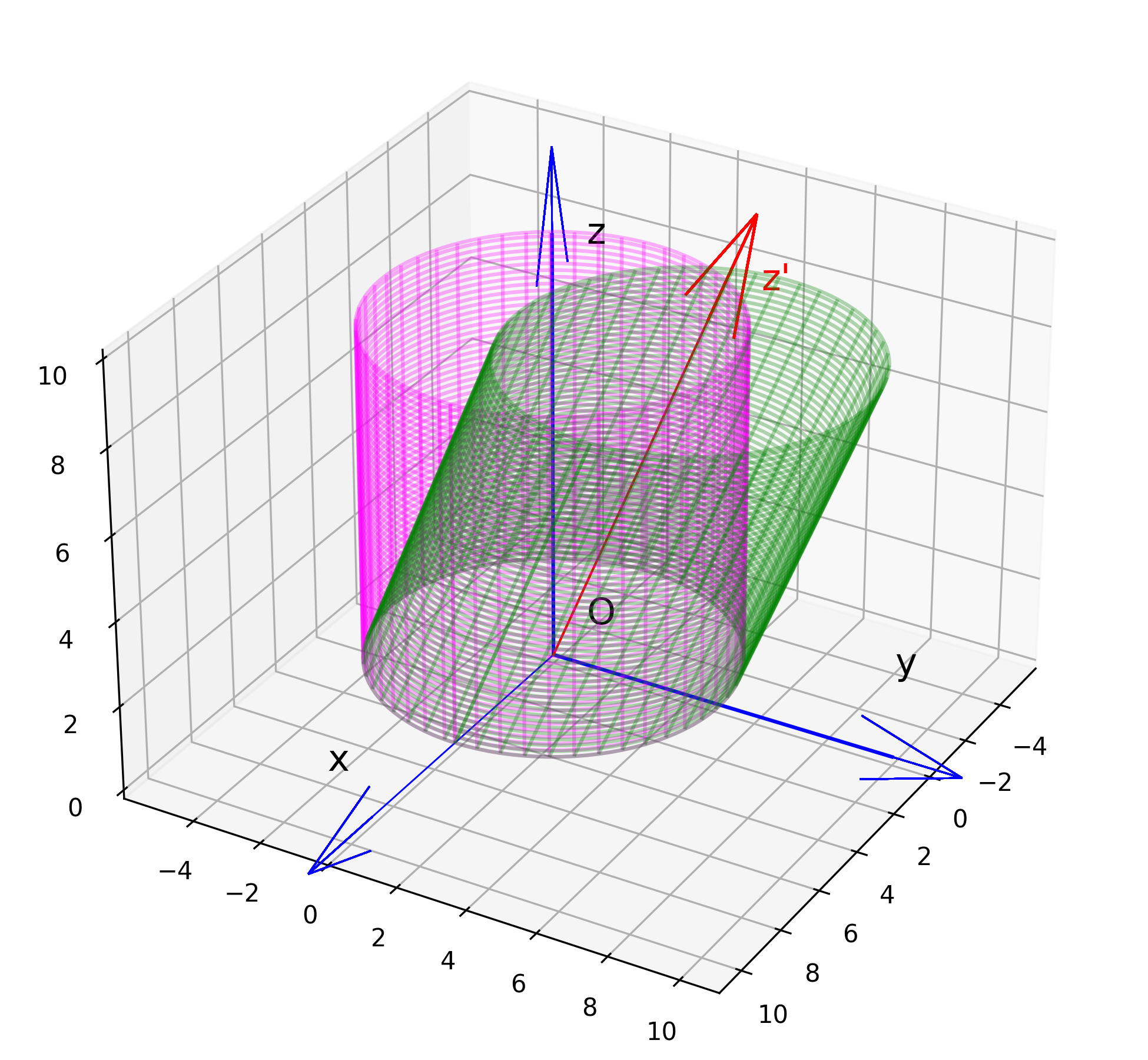}
    \caption{Non-orthogonal z axis shifted cylinder coordinate.}
    \label{fig_shifted_cylinder}
\end{figure}
The above subsection, we have tested the SymFields module in an orthogonal (cylinder) coordinates, now we will benchmark it under non-orthogonal coordinates by given an $\alpha$ angle shift to the original Z coordinate of (orthogonal) cylinder coordinate. As shown in Fig. \ref{fig_shifted_cylinder}, the $z'$ axis is shifted with angle $\alpha$ to z axis in the y-o-z plane. The mapping from original cylinder coordinates to the shifted cylinder coordinates shall be:
\begin{equation}
    \begin{cases}
        x = r'cos\phi' \\
        y = r'sin\phi' + z'sin\alpha \\
        z = z'cos\alpha
    \end{cases}
\end{equation}

With SymFields, we can easily get its covariant metric tensor by:
{\footnotesize
\begin{lstlisting}
# test non-orthogonal shifted cylinder coordinate
r_2, phi_2, z_2, alpha = sympy.symbols('r_2, phi_2, z_2, alpha')
Xi = [r_2, phi_2, z_2]
x = r_2*sympy.cos(phi_2)
y = r_2*sympy.sin(phi_2) + z_2*sympy.sin(alpha)
z = z_2*sympy.cos(alpha)
R = [x, y, z]
M_co = Metric(Xi=Xi, R=R, coordinate='shifted cylinder', contra=0, evaluation=1)
M_co = sympy.simplify(M_co)
M_contra = M_co.inv(method='GE')
\end{lstlisting}
}

{\footnotesize
\begin{lstlisting}
In: M_co
Out: 
\end{lstlisting}
}

\begin{figure}[H]
    \centering
    \includegraphics[height=0.075\linewidth]{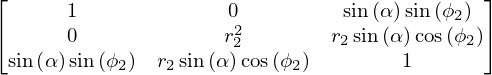}
\end{figure}

Let's further check the curl of gradient and divergence of curl in this non-orthogonal coordinate. To make the expressions more simple, we value the shift angle as: $\alpha=\pi/3$.

* curl of gradient
{\footnotesize
\begin{lstlisting}
alpha = pi/3
Xi = [r_2, phi_2, z_2]
x = r_2*sympy.cos(phi_2)
y = r_2*sympy.sin(phi_2) + z_2*sympy.sin(alpha)
z = z_2*sympy.cos(alpha)
R = [x, y, z]
M_co = Metric(Xi=Xi, R=R, coordinate='shifted cylinder', contra=0, evaluation=1)
M_co = sympy.simplify(M_co)
M_contra = M_co.inv(method='GE')

U2 = sympy.Function('U2')
U2 = U2(r_2, phi_2, z_2)

grad2 = Grad(U2, Xi, coordinate='shifted cylinder', metric=M_contra, evaluation=1)
curl_grad2 = Curl(grad2, Xi, coordinate='shifted cylinder', metric=M_contra, evaluation=1)
\end{lstlisting}
}

{\footnotesize
\begin{lstlisting}
In: grad2
Out: 
\end{lstlisting}
}

\begin{figure}[H]
    \includegraphics[height=0.075\linewidth]{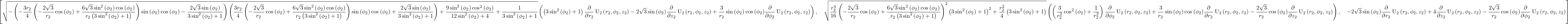}
\end{figure}

From the above result, we can find the analytical expression for the gradient in a non-orthogonal coordinate can be very long and complicated so that it could not be full placed in one line on the page. However, here we only need to verify whether the curl of the gradient under this shifted cylinder coordinate is zero. The result is:

{\footnotesize
\begin{lstlisting}
In: sympy.simplify(curl_grad2[0])
In: sympy.simplify(curl_grad2[1])
In: sympy.simplify(curl_grad2[2])
Out: 
\end{lstlisting}
}

\begin{figure}[H]
    \centering
    \includegraphics[height=0.015\linewidth]{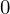}
\end{figure}
We can find that all 3 components of the curl of gradient is 0.

Then let's check the divergence of curl.
{\footnotesize
\begin{lstlisting}
A_r2 = sympy.Function('A_r2')
A_phi2 = sympy.Function('A_phi2')
A_z2 = sympy.Function('A_z2')
A_r2 = A_r2(r_2, phi_2, z_2)
A_phi2 = A_phi2(r_2, phi_2, z_2)
A_z2 = A_z2(r_2, phi_2, z_2)
A2 = [A_r2, A_phi2, A_z2]

curl2 = Curl(A2, Xi, coordinate='shifted cylinder', metric=M_contra, evaluation=1)
div_curl2 = Div(curl2, Xi, coordinate='shifted cylinder', metric=M_contra, evaluation=1)
\end{lstlisting}
}

{\footnotesize
\begin{lstlisting}
In: sympy.simplify(curl2)
Out: 
\end{lstlisting}
}

\begin{figure}[H]
    \centering
    \includegraphics[height=0.085\linewidth]{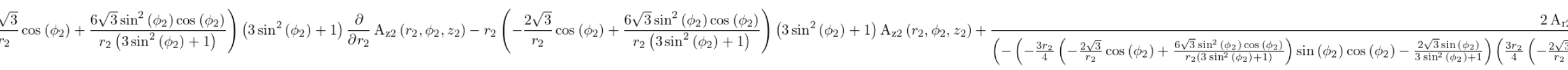}
\end{figure}

We can also find the expression of curl in this $\alpha = \pi/3$ shifted cylinder coordinate is also very long and complicated. However, we can also verify that the divergence of curl remains zero.

{\footnotesize
\begin{lstlisting}
In: sympy.simplify(div_curl2)
Out: 
\end{lstlisting}
}

\begin{figure}[H]
    \centering
    \includegraphics[height=0.015\linewidth]{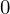}
\end{figure}

\section{Summary}
In this paper, we report the development of an open source symbolic calculation tool for vector field analysis in Python. The SymFields module is constructed upon Python symbolic module sympy, which could only conduct scaler field analysis. Within SymFields module, the vector fields operation are defined upon the metric tensor of a general curvilinear coordinates. Which means you can conduct vector analysis for any curvilinear coordinates regardless whether it is orthogonal or not. Four orthogonal coordinates: Cartesian, Cylinder, Sphere and Toroidal are set at build-in coordinates. You can extend it to any other coordinates by providing a new metric tensor. In SymFields, the differential operators based on metric tensor are normalized to real physical values, which means your can use real physical value of the vector fields as inputs. Thus could greatly free the physicists from the tedious calculation under complicated coordinates.


\begin{thebibliography}{1}

\bibitem{Qin_1999_CPC}
H.~Qin, W.~M. Tang, and G.~Rewoldt.
\newblock Symbolic vector analysis in plasma physics.
\newblock {\em Computer Physics Communications}, 116(1):107 -- 120, 1999.

\bibitem{USTC_2008_book}
\newblock Introduction to Advanced Mathematics (printed in Chinese), University of Science and Technology of China Press, by group author of Mathematics teaching and research team, 2008 Hefei, China.

\bibitem{Piercey_lecture}
V.~I. Piercey.
\newblock Lame and metric coefficients for curvilinear coordinates in {$R^3$}.
\newblock Lecture notes, University of Arizona.

\bibitem{William_1991_book}
William~Denis D’haeseleer, William Nicholas~Guy Hitchon, James~D. Callen, and
  J.~Leon Shohet.
\newblock {\em Flux Coordinates and Magnetic Field Structure}.
\newblock Springer Berlin Heidelberg, 1991.

\bibitem{Lebedev_2010_book}
Leonid~P Lebedev, Michael~J Cloud, and Victor~A Eremeyev.
\newblock {\em Tensor Analysis with Applications in Mechanics}.
\newblock WORLD SCIENTIFIC, 2010.

\end{thebibliography}

\end{document}